\documentclass{article}
\usepackage{spconf,amsmath,graphicx}
\usepackage[urlcolor=blue]{hyperref}
\usepackage{booktabs}

\title{Enhancing Speaking Styles in Conversational Text-to-Speech Synthesis with Graph-based Multi-modal Context Modeling}
\name{
    Jingbei Li$^{1,*}$,\thanks{$*$ Equal contributions.}
    Yi Meng$^{1,*}$,
    Chenyi Li$^1$,
    Zhiyong Wu$^{1,2,\dagger}$,\thanks{$\dagger$ Corresponding author.}
    Helen Meng$^{2}$,
    Chao Weng$^3$,
    Dan Su$^3$
}
\address{
    $^1$ Tsinghua-CUHK Joint Research Center for Media Sciences, Technologies and Systems, \\
    Shenzhen International Graduate School, Tsinghua University, Shenzhen, China\\
    $^2$ Department of Systems Engineering and Engineering Management, \\
    The Chinese University of Hong Kong, Hong Kong SAR, China\\
    $^3$ AI Lab, Tencent, Shenzhen, China\\
    \small{
    \{lijb19,my20,licy20\}@mails.tsinghua.edu.cn, \{zywu,hmmeng\}@se.cuhk.edu.hk, \{cweng,dansu\}@tencent.com
    }
}

\begin{document}
\ninept
\maketitle

\begin{abstract}
Comparing with traditional text-to-speech (TTS) systems, conversational TTS systems are required to synthesize speeches with proper speaking style confirming to the conversational context. However, state-of-the-art context modeling methods in conversational TTS only model the textual information in context with a recurrent neural network (RNN). Such methods have limited ability in modeling the inter-speaker influence in conversations, and also neglect the speaking styles and the intra-speaker inertia inside each speaker. Inspired by DialogueGCN and its superiority in modeling such conversational influences than RNN based approaches, we propose  a graph-based multi-modal context modeling method and adopt it to conversational TTS to enhance the speaking styles of synthesized speeches. Both the textual and speaking style information in the context are extracted and processed by DialogueGCN to model the inter- and intra-speaker influence in conversations. The outputs of DialogueGCN are then summarized by attention mechanism, and converted to the enhanced speaking style for current utterance. An English conversation corpus is collected and annotated for our research and released to public. Experiment results on this corpus demonstrate the effectiveness of our proposed approach, which outperforms the state-of-the-art context modeling method in conversational TTS in both MOS and ABX preference rate.
\end{abstract}

\begin{figure}[t]
  \centering
  \includegraphics[width=\linewidth]{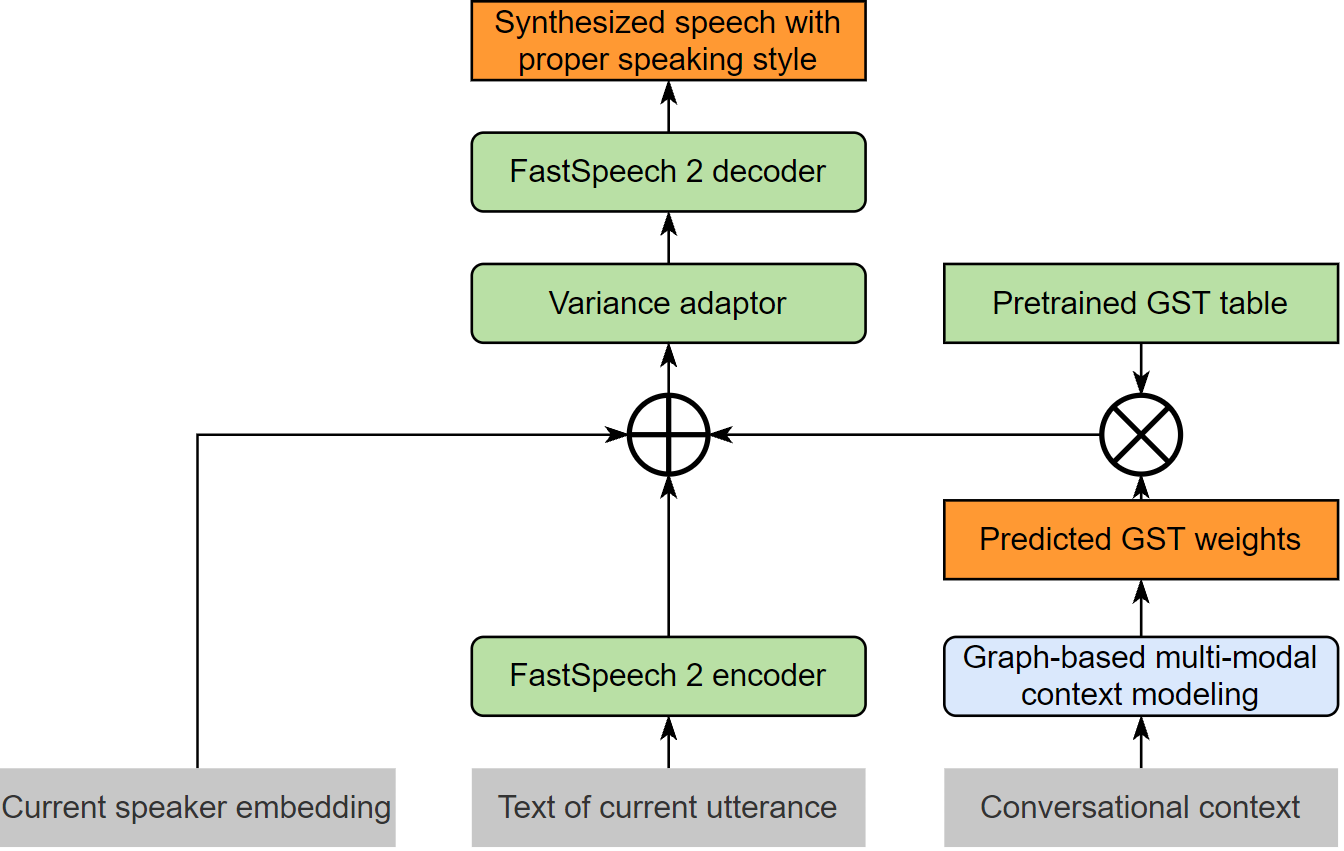}
  \caption{The proposed conversational TTS system. Our work mainly focuses on the graph-based multi-modal context modeling.}
  \label{fig:conversational tts}
\end{figure}

\section{Introduction}

\begin{figure*}[t]
  \centering
  \includegraphics[width=\linewidth]{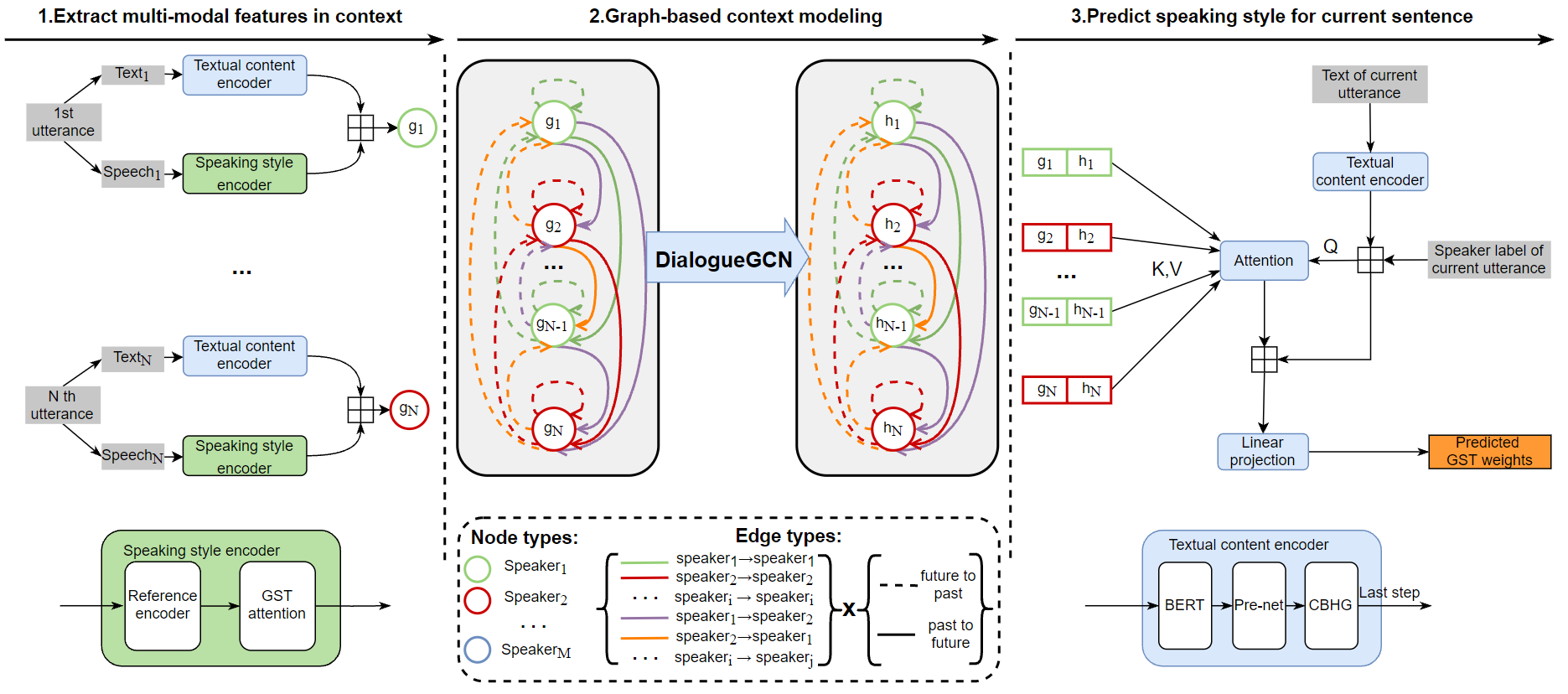}
  \caption{Graph-based multi-modal context modeling}
  \label{fig:dialoguegcn}
\end{figure*}

With the development of deep learning,
neural network based text-to-speech (TTS) systems 
have achieved superior performance
than traditional systems 
\cite{wang_tacotron:_2017, shen_natural_2017, miao_flow-tts_2020}.
Although the current TTS systems have advanced in expressiveness \cite{skerry-ryan_towards_2018, wang_style_2018, lei2021fine, li2021towards},
the TTS systems used in the spoken dialog systems such as virtual assistants and voice agents are still not satisfying due to the inability to generate speech with speaking styles confirming to the conversational context.
In human-human conversations,
people are interacting with different social signals such as humor, empathy and compassion \cite{vinciarelli_open_2015}
through the contents and speaking styles of their speeches.
Moreover, there is always a major interplay between the inter- and intra-speaker dependencies in the conversations \cite{ghosal_dialoguegcn_2019}, where inter-speaker dependency refers to the influence that counterparts produce in a speaker while intra-speaker dependency refers to the inertia inside the speakers themselves.



It is of great importance for the conversational TTS systems to synthesize speeches with proper speech styles in accordance with the conversational context. 
However, state-of-the-art context modeling methods in conversational TTS systems are still in their early stage.
A recent work proposes a conversational context encoder \cite{guo_conversational_2020} to 
model the context
by sequentially processing the textual information 
through a uni-directional 
gated recurrent unit (GRU) \cite{cho_learning_2014} network, 
which however
only deals with textual information,
neglecting the speaking styles in speeches.
Moreover, 
such recurrent neural network (RNN) based method has limited ability in modeling the inter-speaker influence in practise \cite{bradbury2016quasi},
and is hard to model the intra-speaker influence in conversations.



Inspired by the superiority of dialogue graph convolutional network (DialogueGCN) \cite{ghosal_dialoguegcn_2019} in modeling both inter- and intra-speaker influences 
as well as other contexts,
we propose 
a graph-based multi-modal conversational context modeling module
and adopt it to 
conversational TTS 
to enhance the speaking style of synthesized speeches.
Multi-modal features are extracted from the context,
including
textual information and speaking style information of all past utterances in the conversation.
The extracted multi-modal features are then modeled by DialogueGCN 
to produce new representations holding 
richer context influenced by inter- and intra- speaker and temporal dependencies in conversations.
The original multi-modal features and new representations are then summarized by attention mechanism \cite{vaswani_attention_2017}
to predict the speaking style for the current utterance.
Finally the predicted speaking style is fed to 
the proposed conversational TTS system
to synthesize speech of the current utterance with proper speaking style,
which is based on a global style token (GST) \cite{wang_style_2018} enhanced FastSpeech 2 \cite{ren_fastspeech_2020} model.

An English conversation corpus is collected and annotated for the research, and 
released to public.
Experiment results of both objective and subjective evaluations on this corpus
demonstrate the effectiveness of our proposed method over
a baseline approach with state-of-the-art RNN based context modeling method in conversational TTS.
The effectiveness of the uses of the graph-based architecture and the multi-modal information 
are also demonstrated by ablation studies.
\section{Methodology}
\label{methodology}

\subsection{Framework of conversational TTS}
\label{conversational tts}
As shown in Fig.\ref{fig:conversational tts}, 
we employ a GST \cite{wang_style_2018} enhanced FastSpeech 2 \cite{ren_fastspeech_2020} as the backbone of our proposed conversational TTS system.
The proposed graph-based multi-modal context modeling module is adopted to predict the GST weights 
for the current utterance
from the conversational context, which will be elaborated in Section \ref{context modeling}.
The predicted GST weights are then
recovered to 
the speaking style embedding
with the style tokens in the pretrained GST table. 
The encoder output of FastSpeech 2 which contains the textual information 
is added with the speaker embedding and the predicted speaking style embedding
to conclude the necessary information for synthesizing speech of the current utterance.
The sum is processed by the rest modules of FastSpeech 2 to 
to predict the mel-spectrogram.
Finally, a well-trained HiFi-GAN \cite{kong2020hifi} is used as the vocoder to generate speech with desired speaking style confirming to the conversational context.




\subsection{Graph-based multi-modal context modeling}
\label{context modeling}


To model the conversational context,
we employ DialogueGCN \cite{ghosal_dialoguegcn_2019} as the key 
component 
in the proposed graph-based multi-modal context modeling module.

As shown in Fig.\ref{fig:dialoguegcn},
for each past utterance in the conversation, multi-modal features are firstly extracted.
The textual content encoder is utilized to extract textual features from the text of an utterance, which is a pretrained BERT \cite{devlin_bert:_2018} model followed by the Pre-net and CBHG networks in Tacotron \cite{wang_tacotron:_2017}.
The output at the last step of CBHG is used as the textual features.
The speaking style features are generated by passing the acoustic speech of an utterance to a speaking style encoder that is composed of a reference encoder and a GST attention layer \cite{wang_style_2018}.

The past utterances in the conversation are organized as a directed graph.
Each utterance is represented as a node in the graph.
Then a complete directed graph with self-loop edges at each node is generated.
Each edge represents either the inter- or intra-speaker dependency depending on the speaker labels of their corresponding utterances,
along with their relative temporal orders (either \textit{future to past} or \textit{past to future}) in context. 
Particularly, the edge is marked as \textit{future to past} when it is a self-loop edge.

The graph is then initialized with the extracted multi-modal features
and processed by DialogueGCN for 1 iteration
to produce new representations holding 
richer context information influenced by the temporal and speaker dependencies in conversation.


Attention mechanism \cite{vaswani_attention_2017}
is employed to find out the features most related to the text of the current utterance from the original multi-modal features and the new representations from DialogueGCN.
The query is set to the concatenation of textual features extracted by the textual content encoder and the speaker label of the current utterance.
The summarized features (i.e. output of the attention mechanism) are then further concatenated to the query and 
used to predict the GST weights of the current utterance by a linear projection with $softmax$ activation.
The MSE between the predicted and ground-truth GST weights extracted by the speaking style encoder is used as the loss function.

\begin{figure}[tb]
  \centering
  \includegraphics[width=\linewidth]{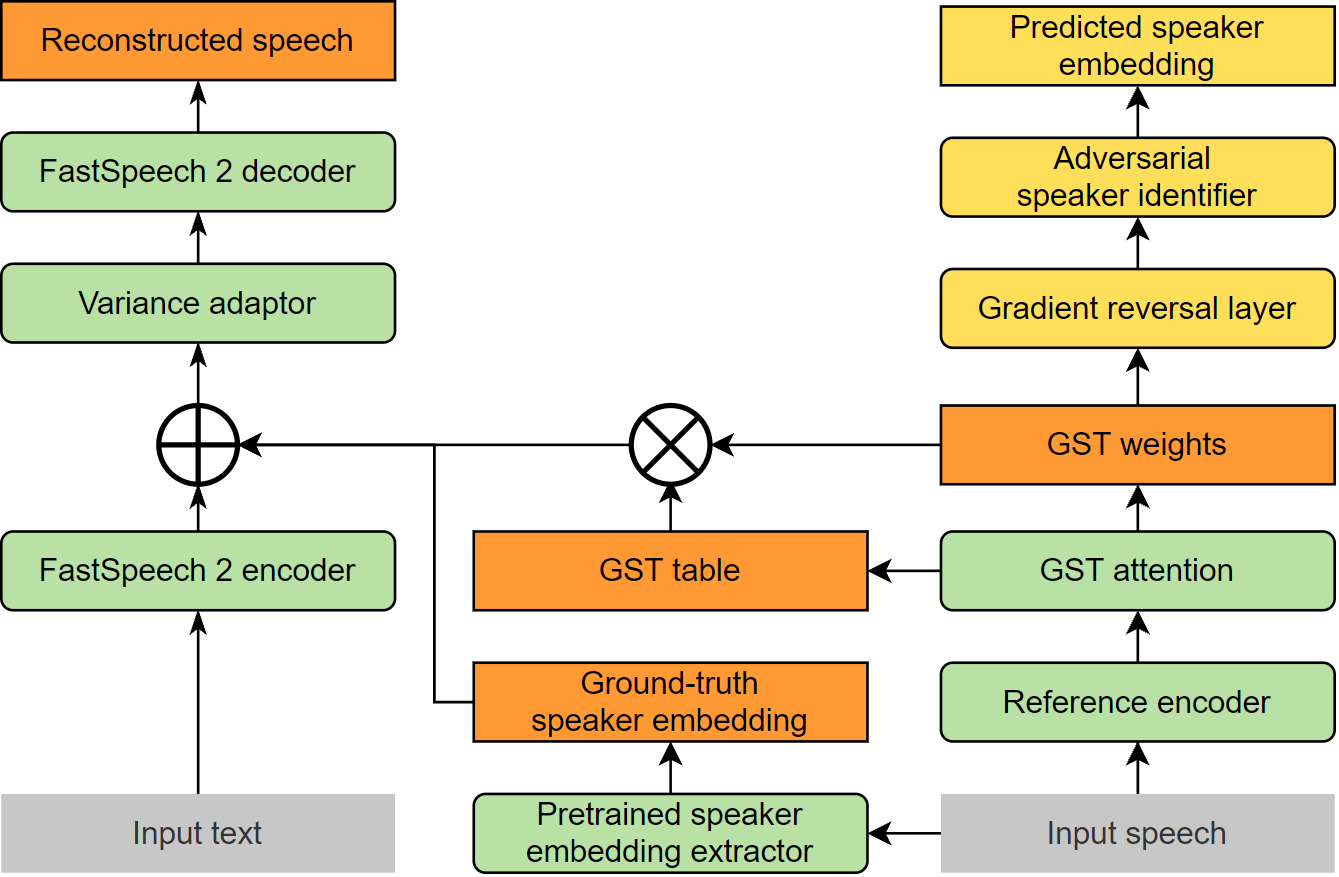}
  \caption{GST enhanced FastSpeech 2}
  \label{fig:gst}
\end{figure}

\subsection{Model pretraining with GST enhanced FastSpeech 2}
\label{gst}

The backbone of our proposed conversational TTS system is a GST enhanced FastSpeech 2, where the 
reference encoder and GST attention layer
are learned in a unsupervised way from conversational corpus.

As shown in Fig.\ref{fig:gst},
the reference encoder and GST attention layer in the original GST learning framework \cite{wang_style_2018} are introduced to extract the speaking style embeddings of the input speech as the weighted sum of the style embeddings in the learnt GST table.
Meanwhile, speaker embedding is extracted from the input speech by a pretrained x-vector \cite{snyder_x-vectors_2018} 
extractor.
The 
speaking style embedding and 
speaker embedding are then added to the encoder output of FastSpeech 2 and further sent to the variance adaptor and decoder to reconstruct the input speech.
Furthermore, 
a speaker adversarial branch \cite{cong_controllable_2021} with gradient reversal layer \cite{ganin_domain-adversarial_2016} is adopted on the learnt GST weights
to disentangle speaker information from the speaking styles.

After the model is pretrained, the reference encoder and the GST attention module are used as the speaking style encoder in our graph-based multi-modal context modeling module, as shown in Fig.\ref{fig:dialoguegcn}.
And the 
rest part
including the encoder, variance adapter, decoder and the learnt GST table 
are transferred to the proposed conversaional TTS framework mentioned in Section \ref{conversational tts}.
\section{English conversation corpus}

We collect 24 hours of speeches from 66 conversational videos from YouTube as the English conversation corpus (ECC) 
\footnote{The English conversation corpus is available at https://github.com/\\thuhcsi/english-conversation-corpus.}
and use it for model training and evaluation in our work.
These videos are originally recorded for conversational English learning for second language learners. 
Therefore all the speakers in ECC have clear and standard English pronunciation.

Each video contains 4 to 68 performed daily conversations. 
And each conversation has 30.4 sentences from 2.9 speakers on average.
The average duration of each sentence is about 2.5s.
In details,
54.9\% of the conversations are performed by 2 speakers, 20.0\% by 3 speakers and 25.0\% by more than 3 speakers.
More statistics of ECC are shown in Table \ref{tab:dataset}.

The transcriptions, speaker labels inside each video, and sentence boundaries are then annotated by a professional annotation group.
Random inspection on the annotations shows that the accuracies of the annotations on the transcriptions and speaker labels have reached 95\%,
and the acceptance rate of sentence boundaries reaches 
95\%.

\begin{table}[t]
  \caption{Statistics of the English conversation corpus (ECC)}
  \label{tab:dataset}
  \centering
  \begin{tabular}{lccc}
    \toprule
    \textbf{For each video} &\textbf{Minium} &\textbf{Average} &\textbf{Maxium} \\
    \midrule
    Number of conversations & 4 & 15.2 & 68 \\
    Number of speakers & 3 & 10.2 & 26 \\
    Duration & 9.1 min & 19.2 min & 38.0 min \\
    \toprule
    \textbf{For each conversation} &\textbf{Minium} &\textbf{Average} &\textbf{Maxium} \\
    \midrule
    Number of sentences & 2 & 30.4 & 222 \\
    Number of speakers & 2 & 2.9 & 9 \\
    Duration & 3.5 s & 75.3 s & 8.8 min \\
    \toprule
    \textbf{For each speaker} &\textbf{Minium} &\textbf{Average} &\textbf{Maxium} \\
    \midrule
    Number of sentences & 1 & 10.5 & 101 \\
    Duration & 0.28 s & 26.1 s & 5.8 min \\
    \toprule
    \textbf{For each sentence} &\textbf{Minium} &\textbf{Average} &\textbf{Maxium} \\
    \midrule
    Number of words & 1 & 5.6 & 49 \\
    Duration & 0.1 s & 2.5 s & 23.1 s \\
    \bottomrule
  \end{tabular}
\end{table}

\section{Experiments}
\label{experiments}

\subsection{Baseline approach}
\begin{figure}[ht]
  \centering
  \includegraphics[width=\linewidth]{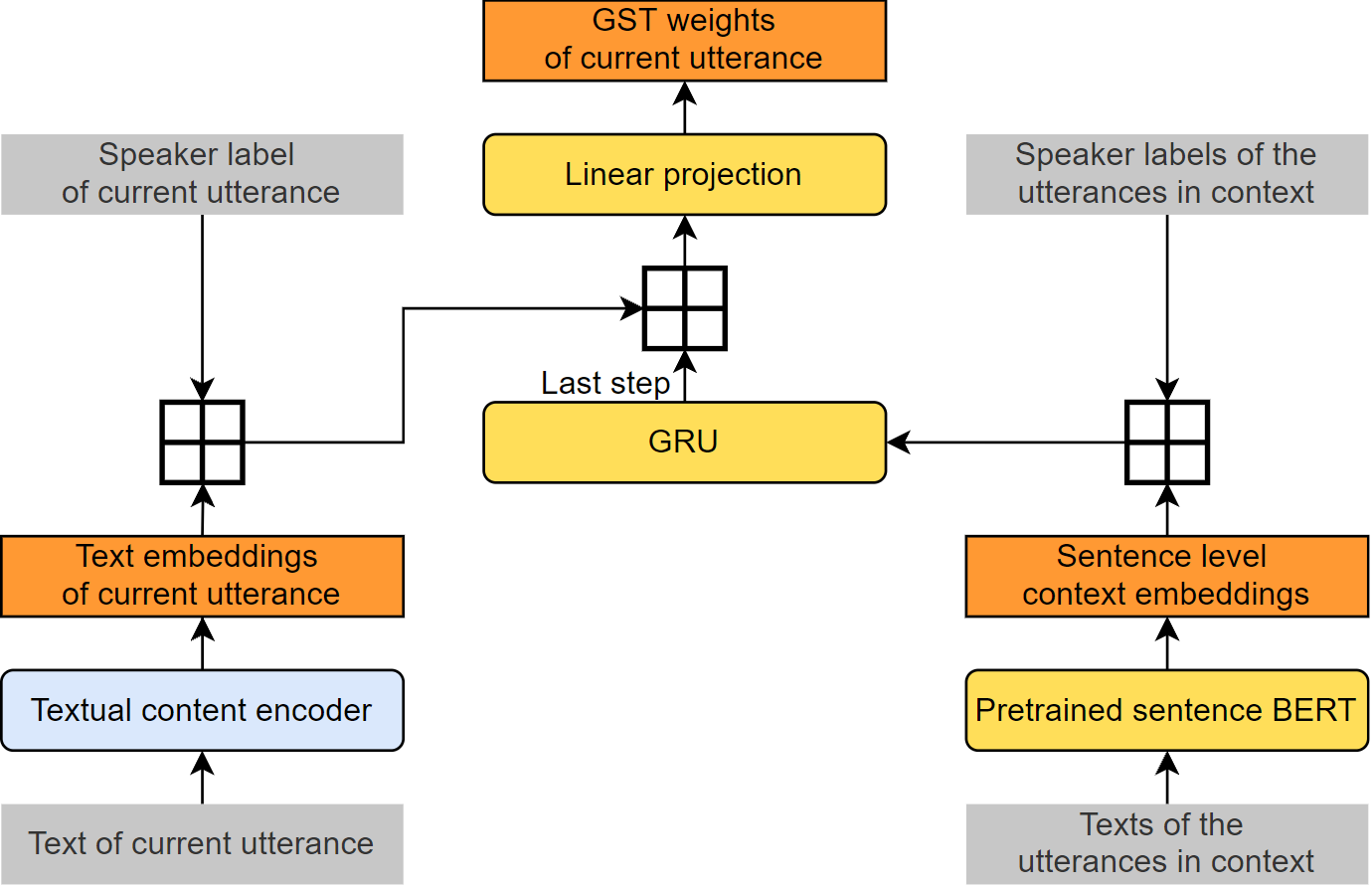}
  \caption{Baseline context modeling method}
  \label{fig:baseline}
\end{figure}

We implement a baseline approach with state-of-the-art context modeling method in conversational TTS \cite{guo_conversational_2020}, as shown in Fig.\ref{fig:baseline}.

The textual features of the current utterance are also extracted by the textual content encoder.
However, the textual features for the utterances in the conversational context are the sentence embeddings extracted by a pre-trained BERT model \cite{devlin_bert:_2018, cui_revisiting_2020}. 
After concatenating with 
their corresponding speaker labels,
the sentence embeddings are modeled by 
a uni-directional GRU with 512 hidden units.
Then the final the state of GRU is concatenated with the textual features and the speaker label of the current utterance and 
used to predict the GST weights for the current utterance by a linear projection with $softmax$ activation.

\subsection{Training setup}
Our training process involves training the GST enhanced FastSpeech 2,
followed by training the graph-based multi-modal context modeling network.

962 conversations from the first 61 videos in ECC corpus are used for training.
28,837 sentences are extracted from these conversations and used to train the GST enhanced FastSpeech 2.
The training setups are the same to those in FastSpeech 2 \cite{ren_fastspeech_2020}. 
The phoneme embedding in the FastSpeech 2 encoder is initialized and fixed as the well-trained parameters learnt on the LJSpeech \cite{ito_lj_2017} corpus.
The model is 
trained 
for 900,000 iterations with a batch size of 16.
And the learning rate is 0.0625 for the first 300,000 iterations 
and 0.02 for the next 400,000 iterations
and 0.005 for the rest iterations. 

To train the proposed speaking style learning model with graph-based multi-modal context modeling,
we convert these conversations into 24,058 conversation chunks, in which
each chunk consists of 6 utterances, 
from which the first 5 utterances 
are used as the past utterances and the conversational context 
for the last one.
The model is trained for 15,000 iterations with a batch size of 32 and a learning rate of $10^{-4}$.

The experiments are implemented with PyTorch \cite{paszke_pytorch_2019} on an NVIDIA 2080 Ti GPU.

\subsection{Evaluations}

1,401 conversation chunks extracted from the rest 5 unseen videos is used for objective evaluation,
in which 25 conversation chunks are further randomly selected for subjective evaluation
\footnote{Samples are available at https://thuhcsi.github.io/icassp2022-\\conversational-tts/.}.

We employ the MSE between the predicted and ground-truth GST weights as the objective evaluation metric.
Speech files generated on the 25 conversation chunks by the baseline and proposed approaches
are rated by 10 listeners on a scale from 1 to 5 with 1 point increments,
from which a subjective mean opinion score (MOS) is calculated.
Meanwhile, the listeners are asked to choose a preferred speech from the speeches synthesized by the baseline and proposed approaches, from which ABX preference rates are calculated.

As shown in Table \ref{tab:objective} and \ref{tab:subjective},
the experimental results
in both objective and subjective evaluations
demonstrate the effectiveness of our proposed graph-based multi-modal context modeling.
The MSE of the predicted GST weights decreases from $5.15 \times 10^{-3}$ to $4.17 \times 10^{-3}$,
MOS score increases from 3.356 to 3.584
and ABX preference rate exceeds the baseline by 32.19\%.
It is also reported by the listeners that the speeches synthesized by the proposed approach have more variable speaking styles such as different speaking rate, emphasis and prosody.

\subsection{Ablation studies}

\subsubsection{Effectiveness of the graph-based context modeling}
We implement a variant of the baseline approach which migrates the graph-based architecture to its context modeling.
In this variant, the sentence embeddings extracted by the pre-trained BERT model
are used as the textual features of the utterances in conversational context for graph-based context modeling.
The speaking style information in context is still not considered.
Objective evaluation of this approach is shown in the second row of Table \ref{tab:objective}.
Comparing with the baseline approach,
the MSE decreases from $5.15 \times 10^{-3}$ to $4.47 \times 10^{-3}$ as expected,
which demonstrates the effectiveness of the proposed graph-based context modeling method.


\subsubsection{Effectiveness of the speaking style information in context}
Based on the proposed approach, we implement another variant which excludes the speaking style information of the past utterances in conversation.
Objective evaluation of this approach is shown in the last row of Table \ref{tab:objective}.
Compared with the proposed approach,
the MSE of the this approach increases to $4.38 \times 10^{-3}$, showing the necessity of using multi-modal information in conversational TTS.
And compared with the previous variant,
the MSE of the this approach decreases by $0.09 \times 10^{-3}$, also showing the superiority of the used textual content encoder than sentence level BERT.
\begin{table}[t]
  \caption{Objective evaluations for different approaches}
  \label{tab:objective}
  \centering
  \begin{tabular}{lc}
    \toprule
    \textbf{Approach}&\textbf{MSE ($\downarrow$)} \\
    \midrule
    \textbf{Baseline} & $5.15 \times 10^{-3}$ \\
     - w/ graph-based context modeling & $4.47 \times 10^{-3}$ \\
    \textbf{Proposed} & $\mathbf{4.17} \times 10^{-3}$ \\
     - w/o speaking style information & $4.38 \times 10^{-3}$ \\
    \bottomrule
  \end{tabular}
\end{table}
\begin{table}[t]
  \caption{Subjective evaluations between the baseline and the proposed approaches. * NP stands for no preference.}
  \label{tab:subjective}
  \centering
  \begin{tabular}{cccc}
    \toprule
    &\textbf{Baseline} &\textbf{NP$^*$} &\textbf{Proposed} \\
    \midrule
    \textbf{MOS} & $3.356 \pm 0.091$ & - & $\mathbf{3.584 \pm 0.100}$ \\
    \textbf{ABX Preference} & $26.22\%$ & $14.67\%$ & $\mathbf{55.11\%}$ \\
    \bottomrule
  \end{tabular}
\end{table}

\section{Conclusions}
\label{conclusions}

To improve the prosody of synthesized speech
in conversaional TTS,
we propose the graph-based multi-modal context modeling
and its application in conversational TTS to enhance speaking styles in synthesized speeches. 
Both the textual and speaking styles in contexts are extracted and modeled by 
DialogGCN to learn the inter- and intra-speaker influences in conversations.
Then the outputs of DialogGCN are summarized by attention mechanism and used to predict the speaking style for current sentence. 
We collect and annotate an English conversation corpus 
to support our research
and is released to public.
Compared with the state-of-the-art context modeling method in conversational TTS
the proposed approach outperforms state-of-the-art approach 
in both objective and subjective evaluations.


\section{Acknowledgements}
\label{acknowledgements}
This work is supported by National Natural Science Foundation of China (NSFC) (62076144, 61433018),
the joint research fund of NSFC-RGC (Research Grant Council of Hong Kong) (61531166002, N$\_$CUHK404$/$15),
the Major Project of National Social Science Foundation of China (NSSF) (13$\&$ZD189).
We would also like to thank 2020 Tencent Rhino-Bird Elite Training Program, Tencent AI Lab Rhino-Bird Focused Research Program (No. JR202041) and Tsinghua University - Tencent Joint Laboratory for the support.
We also thank Qiao Tian, Yuping Wang, Yuxuan Wang from ByteDance for their support in data annotation.

\bibliographystyle{IEEEbib}
\bibliography{references}


\end{document}